\documentclass[aps,superscriptaddress,twocolumn]{revtex4}%
\usepackage{amsfonts}
\usepackage{amsmath}
\usepackage{amssymb}
\usepackage{graphicx}
\usepackage{float}
\begin{document}

\title{Mechanical properties and microdomain separation of fluid membranes with anchored polymers}

\keywords {spontaneous curvature, bending rigidity, edge line
tension, interfacial line tension, microdomain separation}

\author{Hao Wu\ \footnote{E-mail: wade@issp.u-tokyo.ac.jp}}
\affiliation{Institute for Solid State Physics, University of Tokyo,
Kashiwa, Chiba 277-8581, Japan. }
\author{Hayato Shiba}
\affiliation{Institute for Solid State Physics, University of Tokyo,
Kashiwa, Chiba 277-8581, Japan. }
\author{Hiroshi Noguchi\ \footnote{E-mail: noguchi@issp.u-tokyo.ac.jp}}
\affiliation{Institute for Solid State Physics, University of Tokyo,
Kashiwa, Chiba 277-8581, Japan. }

%\pacs{87.16.D-,87.17.Aa,82.70.Uv}

\begin{abstract}
The entropic effects of anchored polymers on biomembranes are
studied using simulations of a meshless membrane model combined with
anchored linear polymer chains. The bending rigidity and spontaneous
curvature are investigated for anchored ideal and excluded-volume
polymer chains. Our results agree with the previous theoretical
predictions well. It is found that the polymer reduces the line
tension of membrane edges, as well as the interfacial line tension
between membrane domains, leading to microdomain formation. Instead
of the mixing of two phases as seen in typical binary fluids,
densely  anchored polymers stabilize small domains. A mean field
theory is proposed for the edge line tension reduced by  anchored
ideal chains, which well reproduces our simulation results.

\end{abstract}
\maketitle

\section{Introduction}

Our knowledge of the heterogeneous structure of biomembranes has
advanced from the primitive fluid mosaic model \cite{Singer72} to
the modern raft model \cite{Simons97,Ikonen01} over the past
decades. According to this modern model, membrane proteins are not
randomly distributed in lipid membranes but concentrated in local
microdomains, called lipid rafts, with a diameter of $10\sim100$
nm~\cite{Vereb03,Korade,Pike}. The raft contains high concentrations
of glycosphingolipids and cholesterol, and plays important roles in
many intra- and intercellular processes including signal transaction
and membrane protein trafficking.

In the last decade, phase separation in multi-component lipid
membranes has been intensively investigated in three-component
systems of saturated and unsaturated phospholipids and
cholesterol~\cite{hone09,lipo03,baga09,Keller02,Tobias03,Hammond05,Kuzmin05,yana08,Schick09}.
Lipid domains exhibit various interesting patterns on the micrometer
scale, which can be reproduced by theoretical calculations and
simulations. Various shapes of lipid domains can be also formed in
the air--water interface \cite{mcco88,HaoWu09,iwam10}. However, the
formation mechanism of microdomains on the $10\sim100$ nm scale has
not been understood so far. In lipid rafts, glycolipids contain
glycan chains. Recently, network-shaped domains and small scattered
domains have been observed in lipid membranes with PEG-conjugated
cholesterol \cite{Miho12}. The effects of  anchored polymers have
been well investigated in the case of uniform  anchoring on
membranes, but the effects on the lipid domains and line tension
have not been well investigated. In this paper, we focus on the
effects of  anchored polymers on the properties of biomembranes, in
particular, on lipid domains.

It is well known that  anchored polymers modify membrane properties.
 The polymer  anchoring induces a positive spontaneous
curvature $C_0$ of the membranes and increases the bending rigidity
$\kappa$. These relations are  analytically derived using the
Green's function method for mushroom region \cite{Lip95,Lip96} and
confirmed by Monte Carlo
simulations~\cite{brei00,Auth03,Auth05,wern10}.
 The membrane properties in  brush region are analyzed
by scaling method~\cite{Lip95,Lip96,Marsh03}. Experimentally, the
$\kappa$ increase is measured by micropipette aspiration of
liposomes~\cite{Evans97}. Polymer decoration can enhance the
stability of lipid membranes. PEG-conjugated lipids can reduce
protein adsorption and adhesion on cellular surfaces, whereby
 PEG-coated liposomes can be used as a drug carrier in drug-delivery
systems \cite{lasi94,hoff08}.

When vesicles are formed from the self-assembly of surfactant
molecules via micelle growth, the vesicle size is determined
kinetically by the competition between the bending energy and the
line tension energy of the membrane
edge~\cite{helf74,from83,kale89,weis05,leng02,made11,brys05,nogu06a,nogu13}.
Recently, Bressel {\it et al.} reported that the addition of an
amphiphilic copolymer induces the formation of larger vesicles~\cite{bres12}. A polymer-anchoring-induced liposome-to-micelle
transition is also observed~\cite{Marsh03,scha02,john03}. The line
tension of the membrane edge is considered to be reduced by polymer
anchoring, but it has not been systematically investigated so far.
In this study, we simulate the
 edge line tension for anchored ideal and excluded-volume
chains and analytically investigate the polymer effects on the edge
tension for ideal chains.

In order to simulate the polymer-anchoring effects on biomembranes,
we employ  one of the solvent-free meshless membrane models
\cite{nog09,nog06}. Since we focus on the entropic effects of
polymer chains, the detailed structures of the bilayer can be
neglected, and thus the membranes can be treated as a curved
surface. In the meshless model, a membrane particle represents a
patch of bilayer membrane and membrane properties can be easily
controlled.

In Sec. II, the membrane model and simulation method are described.
In Sec. III, the bending rigidity and spontaneous curvature are
estimated from the axial force measurement of cylindrical membranes
and are also compared with the previous theoretical predictions. The
reduction in the edge tension is discussed for both ideal chains and
excluded-volume chains in Sec IV. In Sec V, we present our
investigation how  polymer anchoring modifies membrane phase
separation. Finally, a summary and discussion are provided in Sec
VI.

\section{Model and method}

In this study, we employ a coarse-grained meshless membrane model
\cite{nog06} with anchored linear polymer chains. One membrane
particle  possesses only translational degrees of freedom. The
membrane particles form a quasi-two-dimensional (2D) membrane
according to a curvature potential based on the moving least-squares
(MLS) method \cite{nog06}. Polymer particles are linked by a
harmonic potential, and freely move as a self-avoiding chain with a
soft-core repulsion. One end of each polymer chain is anchored on a
single membrane particle with a harmonic potential and a soft-core
repulsion \cite{HaoWu13}. First, we simulate a single-phase
membrane, where all membrane particles including polymer-anchored
particles are the same type (A) of membrane particles. Then, we
investigate membrane phase separation, where a membrane consists of
two types (A and B)  of membrane particles. The polymers are
anchored to the type A particles.

We consider a single- or multi-component membrane composed of
$N_{\rm {mb}}$ membrane particles. Among them, $N_{\rm {chain}}$
membrane particles are anchored by polymer chains. Each polymer
chain consists of $N_{\rm {p}}$ polymer beads with an anchored
membrane particle. The membrane and polymer particles interact with
each other via a potential
\begin{equation}
U_{\rm {tot}}=  U_{\rm {rep}} + U_{\rm {mb}} +  U_{\rm {p}} + U_{\rm
{AB}}, \label{totpot}
\end{equation}
where $U_{\rm {rep}}$ is an excluded-volume potential, $U_{\rm
{mb}}$ is a membrane potential, $U_{\rm {p}}$ is a polymer
potential,
 and $U_{\rm {AB}}$ is a repulsive potential between different species of membrane particles.

All particles have a soft-core excluded-volume potential with a
diameter of $\sigma$.
\begin{equation}
\frac{U_{\rm {rep}}}{k_{\rm
B}T}=\varepsilon\sum_{i<j}\exp\left[-20(r_{ij}/\sigma-1)
+B\right]f_{\rm {cut}}(r_{ij}/\sigma), \label{rep}
\end{equation}
in which $k_{\rm B}T$ is the thermal energy and $r_{ij}$ is the
distance between membrane (or polymer) particles $i$ and $j$. The
diameter $\sigma$ is used as the length unit, $B=0.126$, and $f_{\rm
{cut}}(s)$ is a $C^{\infty}$ cutoff function
\begin{equation}
f_{\rm {cut}}(s)=\left\{
\begin{array}{ll}
\exp\left\{A\left[1+\frac{1}{(|s|/s_{\rm
{cut}})^{n}-1}\right]\right\} &(s<s_{\rm {cut}})
 \\
0 &(s\geqslant s_{\rm {cut}})
\end{array}
\label{cutoff} \right.
\end{equation}
with $n=12$, $A=1$ and $s_{\rm {cut}}=1.2$.

For excluded-volume polymer chains, all pairs of particles including
pairs of polymer beads have the repulsive interaction given in
Eq.~\eqref{rep}. In contrast, for ideal polymer chains, polymer
beads have the excluded-volume interactions only with membrane
particles to prevent polymer beads from passing through the
membrane.

\subsection{Meshless membrane model}

The membrane potential $U_{\rm {mb}}$ consists of attractive and
curvature potentials,
\begin{equation}
\frac{U_{\rm {mb}}}{k_{\rm B}T}=
 \sum_i^{N_{\rm {mb}}} \big[ \varepsilon U_{\rm {att}}(\rho_i)
+k_{\alpha}\alpha_{\rm {pl}}({\mathbf{r}}_{i}) \big],
\label{mempotential}
\end{equation}
where the summation is taken only over the membrane particles. The
attractive multibody potential is employed to mimic the
``hydrophobic'' interaction.
\begin{equation}
U_{\rm {att}}(\rho_i)=
0.25\ln\left\{1+\exp\left[-4(\rho_{i}-\rho^{\ast})\right]\right\}-C,
\label{att}
\end{equation}
which is a function of the local density of membrane particles
$\rho_{i}=\sum_{j\neq i}^{N_{\rm {mb}}} f_{\rm
{cut}}(r_{ij}/\sigma)$, with $s_{\rm {half}}=1.8$ and $s_{\rm
{cut}}=s_{\rm {half}}+0.3$, where $f_{\rm {cut}}(s_{\rm
{half}})=0.5$, which implies $A=\ln(2)\{(s_{\rm {cut}}/s_{\rm
{half}})^{12}-1\}$. The constant $C$ is set for $U_{\rm
{att}}(0)=0$. Here we use $\rho^{\ast}=6$ in order to simulate a 2D
fluid membrane.
 For $\rho_{i} \lesssim \rho^{\ast}$, $U_{\rm {att}}$ acts as a pairwise potential
with $U_{\rm {att}}=-2\sum_{j> i}f_{\rm {cut}}(r_{ij}/\sigma)$, and
for $\rho_{i}\gtrsim \rho^{\ast}$, this potential saturates to the
constant $-C$.

The curvature potential is expressed by the
 shape parameter called ``aplanarity'', which is defined by
\begin{equation}
\alpha_{\rm {pl}}=\frac{9D_{\rm {w}}}{T_{\rm w}M_{\rm w}} ,
\label{apla}
\end{equation}
with the determinant $D_{\rm w}=\lambda_1\lambda_2\lambda_3$, the
trace $T_{\rm w}=\lambda_1+\lambda_2+\lambda_3$, and the sum of the
principal minors $M_{\rm
w}=\lambda_1\lambda_2+\lambda_2\lambda_3+\lambda_3\lambda_1$. The
aplanarity $\alpha_{\rm {pl}}$ scales the degree of deviation from
the planar shape, and $\lambda_1$, $\lambda_2$, and $\lambda_3$ are
three eigenvalues of the weighted gyration tensor
\begin{equation}
a_{\alpha\beta}({\mathbf{r}}_{i})=\sum_j^{N_{\rm {mb}}}
(\alpha_j-\alpha_G)(\beta_j-\beta_G)w_{\rm{cv}}(r_{ij}),
\label{eigen}
\end{equation}
where $\alpha,\beta\in \left\{x,y,z\right\}$, and the mass center of
a local region of the membrane
$\mathbf{r}_G=\sum_j\mathbf{r}_jw_{\rm{cv}}(r_{ij})/\sum_j
w_{\rm{cv}}(r_{ij})$. A truncated Gaussian function is employed to
calculate the weight of the gyration tensor
\begin{equation}
w_{\rm{cv}}(r_{ij})=\left\{
\begin{array}{ll}
\exp\left[\frac{(r_{ij}/r_{\rm {ga}})^2}{(r_{ij}/r_{\rm
{cc}})^{12}-1}\right] &(r_{ij}<r_{cc})
 \\
0 &(r_{ij}\geqslant r_{\rm {cc}}),
\end{array}
\right. \label{weight}
\end{equation}
which is smoothly cut off at $r_{ij}=r_{\rm {cc}}$. Here we use the
parameters $r_{\rm {ga}}=0.5r_{\rm {cc}}$ and $r_{\rm
{cc}}=3\sigma$. The bending rigidity and the edge tension
 are linearly dependent on $k_{\alpha}$ and
$\varepsilon$, respectively, for $k_{\alpha} \gtrsim 10$, so that
they can be independently varied by changing $k_{\alpha}$ and
$\varepsilon$, respectively.

\subsection{Anchored Polymer Chain}

We consider flexible linear polymer chains anchored on the membrane.
Polymer particles are connected by a harmonic spring potential,
\begin{equation}
\frac{U_{\rm {p}}}{k_{\rm B}T}= \frac{k_{\rm {bond}}}{2} \sum_{\rm
chains} (r_{i,i+1}-b)^2, \label{mempolypoten}
\end{equation}
where $k_{\rm {bond}}$ is the spring constant for the harmonic
potential and  $b=1.2\sigma$ is the Kuhn length of the polymer
chain. The summation is taken only between neighboring particles
along polymer chains and between the end polymer particles and
anchored membrane particles (a total of $N_{\rm p}$ springs in each
chain).

\subsection{Phase separation in membranes}

Two types of membrane particles, A and B, are considered in Sec.
\ref{sec:dom}. The number of these particles are $N_{\rm A}$ and
$N_{\rm B}$, respectively. To investigate phase separation, we apply
a repulsive term $U_{\rm {AB}}$ in Eq.~\eqref{totpot} to reduce the
chemical affinity between different types of membrane particles.
\cite{nog12} The potential $U_{\rm {AB}}$ is a monotonic decreasing
function: $U_{\rm {AB}}/k_{\rm B}T= \varepsilon_{\rm {AB}} \sum_{i
\in A, j\in B} A_1f_{\rm {cut}}(r_{i,j})$ with $n=1$, $A_1=1$ and
$r_{\rm {cut}}=2.1\sigma$, and $A_1=\exp[\sigma/(r_{\rm
{cut}}-\sigma)]$ to set $U_{\rm {AB}}(\sigma)=1$.

\subsection{Simulation method}

The $NVT$ ensemble (constant number of particles $N$, volume $V$,
and temperature $T$) is used with periodic boundary conditions in a
simulation box of dimensions $L_x\times L_y\times L_z$. For planar
membranes, the projected area $L_x\times L_y$ is set for the
tensionless state. The dynamics of both membrane and anchored
flexible polymers are calculated by using underdamped Langevin
dynamics. The motions of membrane and polymer particles are
governed by
\begin{equation}
m\frac{d^2\mathbf{r}_{i}}{dt^2}=-\frac{\partial U_{\rm
{tot}}}{\partial
\mathbf{r}_{i}}-\zeta\frac{d\mathbf{r}_{i}}{dt}+\mathbf{g}_{i}(t),
\label{Lang1}
\end{equation}
where $m$ is the mass of a particle (membrane or polymer particle)
and $\zeta$ is the friction constant. $\mathbf{g}_{i}(t)$ is a
Gaussian white noise, which obeys the fluctuation-dissipation
theorem. We employ the time unit $\tau=\zeta\sigma^2/k_{\rm B}T$
with $m= \zeta\tau$. The Langevin equations are integrated by the
leapfrog algorithm \cite{alle87} with a time step of $\Delta
t=0.005\tau$.

\begin{figure}
  \includegraphics[width=8.5cm]{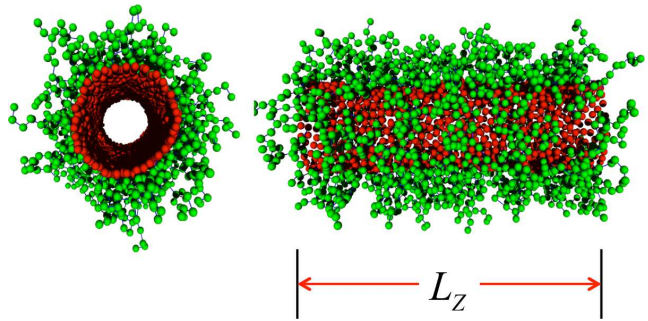}
  \caption{\label{fig:snapcyl}
Front and side views of a snapshot of a cylindrical membrane with
anchored excluded-volume polymer chains at the polymer density
$\phi=0.167$ and cylinder axial length $L_z= 45.3\sigma$. The red
and green particles represent membrane and polymer particles,
respectively. }
\end{figure}

We use $N_{\rm p}=10$, $\varepsilon=4$, $k_{\alpha}=10$, and $k_{\rm
{bond}}=10$ throughout this study. In the absence of anchored
polymer chains, the tensionless membranes have a bending rigidity of
$\kappa/k_{\rm B}T= 21 \pm 0.5$, the edge tension
 $\Gamma_{\rm {ed}}\sigma/k_{\rm B}T=4.5$ and the area
$a_0=1.44\sigma^2$ per membrane particle \cite{Hay11}.  For the
single-phase membranes, the number of membrane particles is fixed as
$N_{\rm {mb}}=1200$ and the number fraction $\phi=N_{\rm
{chain}}/N_{\rm {mb}}$ of polymer-anchored membrane particles is
varied.  To investigate phase separation, the number of the type A
membrane particles is fixed as $N_{\rm {A}}=400$, and the number of
the type B particles is chosen as $N_{\rm {B}}=400$ and $2100$ for a
striped domain and a circular domain, respectively. The polymer
chains are anchored to the type A particles and the polymer fraction
$\phi=N_{\rm {chain}}/N_{\rm {A}}$ is varied. To confirm that the
membranes are in thermal equilibrium, we compare the results between
two initial states, stretching or shrinking, and check that no
hysteresis has occurred. We slowly stretch and shrink cylindrical or
striped membranes in the axial direction  with a speed less than
$dL_z/dt=10^{-6}\sigma/\tau$
 and then equilibrate them for $t/\tau=6\times 10^4$ before the measurements.
For the simulations of circular domains, the membranes were
equilibriated for a duration of $6\times 10^{4}\tau$ after step-wise
changes of $\varepsilon_{\rm {AB}}$. The error bars are calculated
from six independent runs.

\section{Bending rigidity and spontaneous curvature of membranes}\label{sec:scv}

\begin{figure}
\includegraphics[width=8.cm]{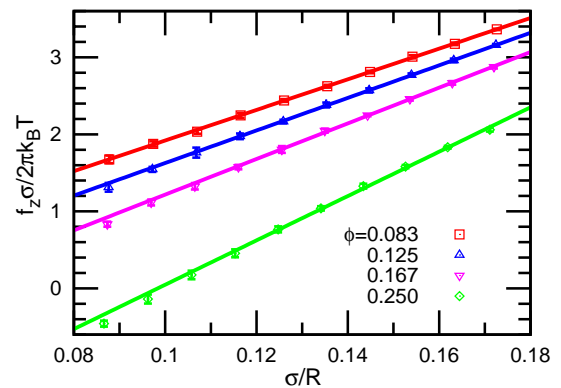}
  \caption{\label{fig:fz}
Force $f_z$ dependence on the radius $R$ of the cylindrical
membranes with anchored excluded-volume chains at $\phi=0.083$,
$0.125$, $0.167$, and $0.250$. The solid lines are obtained by
linear least-squares fits.}
\end{figure}

\begin{figure}
\includegraphics[width=8.cm]{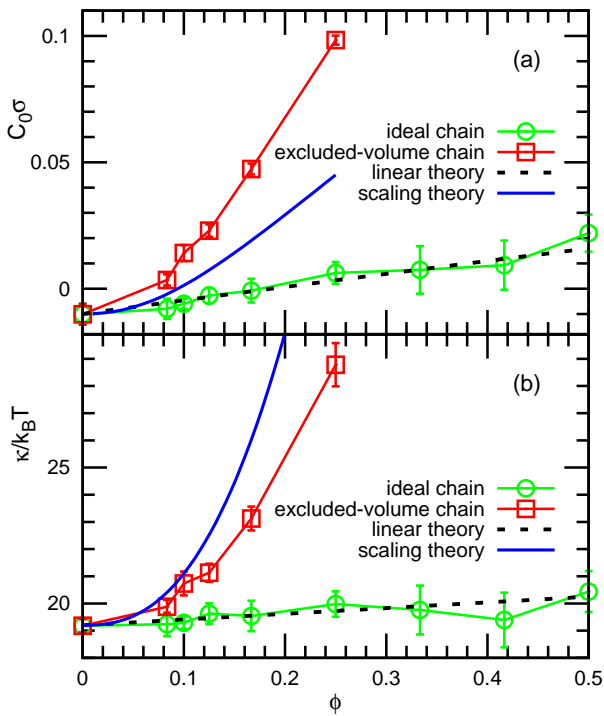}
  \caption{\label{fig:spcv}
Polymer density  $\phi$ dependence of (a) the spontaneous curvature
$C_0$ and (b) bending rigidity $\kappa$ of the membranes with
anchored ideal chains and excluded-volume chains. The dashed lines
in (a) and (b) represent the prediction of the linear theory
(Eqs.~\eqref{eq:scvp} and \eqref{eq:kappap}) for the ideal chains.
The solid lines in (a) and (b) represent the prediction of the
scaling theory (Eqs.~\eqref{eq:brushc} and \eqref{eq:brushk} ) in
the brush region. }
\end{figure}

A cylindrical membrane with polymer chains anchored outside the
membrane is used to estimate the polymer-induced spontaneous
curvature and bending rigidity (see Fig. \ref{fig:snapcyl}). For a
cylindrical membrane with radius $R$ and length $L_z$, the Helfrich
curvature elastic free energy is given by
\begin{eqnarray}
F_{\rm {cv}} &=& \int \Big[ \frac{\kappa}{2} (C_1 + C_2- C_0)^2 +
\bar{\kappa} C_1C_2 \Big] dA \nonumber \\ \label{eq:cy}
 &=& 2\pi RL_z \left[ \frac{\kappa}{2}\left( \frac{1}{R} -C_0 \right)^2 \right],
\end{eqnarray}
where $C_1$ and $C_2$ are the principal curvatures at each position
on the membrane surface, and the membrane area $A=2\pi RL_z$. The
coefficients $\kappa$ and $\bar{\kappa}$ are the bending rigidity
and the saddle-splay modulus, respectively, and $C_0$ is the
spontaneous curvature. In the absence of the anchored polymers (we
call it a pure membrane hereinafter), the membrane has zero
spontaneous curvature, $C_0=0$.

The membrane also has an area  compression energy $F_{\rm {ar}}(A)$:
 $F_{\rm {ar}}(A)=K_A(A -A_0)^2/2A_0$ for $A -A_0\ll A_0$,
where $A_0$ is the area of the tensionless membrane. The radius $R$
is determined by free-energy minimization $\partial F/\partial
R|_{L_z}=0$  for $F=F_{\rm {cv}}+F_{\rm {ar}}(A)$. Since the
curvature energy increases with increasing $L_z$, a contractile
force,
\begin{equation}
f_z = \frac{\partial F}{\partial L_z}\Big|_{R} = 2\pi\kappa
\Big(\frac{1}{R} - C_0\Big), \label{eq:fz}
\end{equation}
is generated along the cylindrical axis.

Figure~\ref{fig:fz} shows the axial force $f_z$ calculated from the
pressure tensors,
 \begin{equation}
\label{eq:pressure_tensor} P_{\alpha\alpha} = (Nk_{\rm B}T -
     \sum_{i} \alpha_{i}\frac{\partial U}{\partial {\alpha}_{i}} )/V,
\end{equation}
for $\alpha \in \{x,y,z\}$, where the summation is taken over all
membrane and polymer particles. When the potential interaction
crosses the periodic boundary,
 the periodic image $\alpha_i + nL_\alpha$
 is employed for $P_{\alpha\alpha}$ calculation.
The force $f_z$ increases linearly with $1/R$ \cite{HaoWu13}. Thus,
$C_0$ and $\kappa$ of the  anchored membranes can be estimated from
a linear fitting method to Eq. ~\eqref{eq:fz}. For both anchored
ideal chains and excluded-volume chains, the obtained values of
$C_0$ and $\kappa$ are shown in Fig.~\ref{fig:spcv}. For the pure
membranes, the value of $\kappa$ agrees very well with those
estimated from the height fluctuations of planar
membranes~\cite{Hay11} and membrane buckling \cite{nogu11a}. The
estimated value of $C_0$ for the pure membrane deviates slightly
from the exact value, zero. This small deviation would be caused by
a higher-order term of the curvature energy \cite{harm06} or finite
size effects as discussed in Ref. \cite{Hay11}.

The anchored polymer generates a positive  spontaneous curvature,
and enhances the bending rigidity $\kappa$. Both quantities increase
with increasing polymer chain density, and for the excluded-volume
chains, these increases are enhanced by the repulsive interactions
between the neighboring chains.

In the mushroom region, the spontaneous curvature and  bending
rigidity are linearly dependent on the polymer density $\phi$.
Analytically, the relations \cite{Lip95,Lip96}
\begin{eqnarray}
\kappa \Delta C_0&=&2a_{\rm {sp}}k_{\rm B}T R_{\rm {end}} \phi/a_0, \label{eq:scvp} \\
\Delta \kappa &=&   a_{\kappa}k_{\rm B}T R_{\rm {end}}^2 \phi/a_0,
\label{eq:kappap}
\end{eqnarray}
are predicted, where $\Delta C_0$ and $\Delta \kappa$ are the
differences of the spontaneous curvatures and  bending rigidities
between the polymer-decorated membrane and the pure membrane,
respectively, and
 $R_{\rm {end}}$ is the mean end-to-end distance of the polymer chain. The
factor $2$ in Eq.~\eqref{eq:scvp} appears because in our definition
the spontaneous curvature is twice as large as that in the previous
works \cite{Lip95,Lip96,brei00,Auth03}. The coefficients are derived
analytically using the Green's function \cite{Lip95,Lip96} and also
 estimated by Monte Carlo simulations of single anchored polymer chains~\cite{Auth03}:
 $a_{\kappa}=0.21$ and $0.2$; and $a_{\rm {sp}}=0.18$ and $0.17$
for ideal and excluded-volume chains, respectively. Our results for
the ideal chains agree very well with these previous predictions
(compare the dashed lines and symbols in Fig. \ref{fig:spcv}). To
draw the dashed line in Fig. \ref{fig:spcv}(a), the end-to-end
distance is estimated from the simulation; $R_{\rm
{end}}=4.16\sigma$, which is slightly larger than a free polymer
chain $R_{\rm {end}}=\sqrt{N_{\rm p}}b=3.79\sigma$. Note that
anchored ideal polymer chains are in the mushroom region
 for any density $\phi$,
since the polymer chains do not directly interact with each other.

For excluded-volume chains, our results deviate from the theoretical
predictions (Eq.~\eqref{eq:scvp}) for the mushroom region at $\phi
\gtrsim 0.1$. Thus, in the high density of anchored polymer chains,
the interactions between polymer chains are not negligible. We
compare our results with a scaling theory based on a blob picture
for the brush region in Ref.~\cite{Lip96}.
 For a cylindrical membrane, the spontaneous curvature is derived from the free-energy minimization with respect to the radius of the cylinder, \cite{Lip96}
\begin{equation}
\frac{\partial f_{\rm c}(x)}{\partial x}+\frac{4\kappa_0}{k_{\rm
B}T}N_{\rm p}^{-3}\bar{\Gamma}^{-3/2\nu}x = 0, \label{eq:brushc}
\end{equation}
where $f_{\rm c}(x)=[\{1+(1+\nu)x/\nu\}^{\nu/(1+\nu)}-1]/x-1$, the
bending rigidity $\kappa_0$ of the pure membrane, and the reduced
spontaneous curvature $x=h_0 C_0/2$ for the height of a brush on a
flat surface $h_0=N_{\rm p}\bar{\Gamma}^{(1-\nu)/2\nu}b$. The
polymer coverage on the membrane is normalized by the maximum
coverage as $\bar{\Gamma}=\Gamma/\Gamma_{\rm {max}}=b^2\phi/a_0$,
and the exponent $\nu=0.6$ is used for excluded-volume chains. The
bending rigidity is given by
\begin{equation}
\Delta\kappa= \frac{\nu+2}{12\nu^2}N_{\rm
p}^3\bar{\Gamma}^{3/2\nu}k_{\rm B}T, \label{eq:brushk}
\end{equation}
in the small curvature limit. \cite{Lip96} Our results qualitatively
agree with these predictions from the scaling theory (see
Fig.~\ref{fig:spcv}). The deviation is likely due to the polymer
length ($N_{\rm p}=10$) in the simulation, which is too short to
apply the blob picture in the scaling theory.

\section{Edge line tension}

\begin{figure}
  \includegraphics[width=8.5cm]{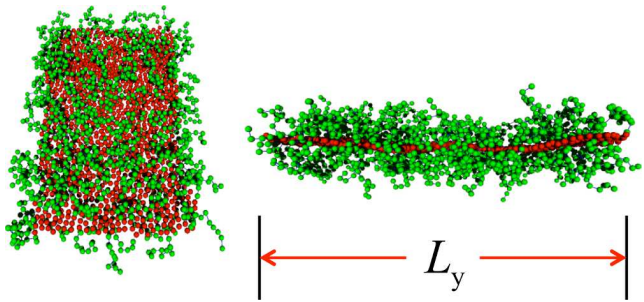}
  \caption{\label{fig:snapstrip}  Top and side views of a snapshot
of a membrane strip with  anchored excluded-volume chains at
$\phi=0.15$ and the length $L_y= 57.6\sigma$ of each membrane edge.
}
\end{figure}

\subsection{Simulation results}

\begin{figure}
  \includegraphics[width=8.cm]{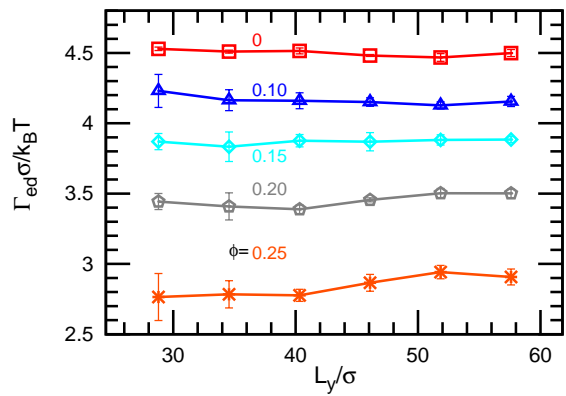}
  \caption{\label{fig:nochange}  Edge line tension $\Gamma_{\rm {ed}}$
of a membrane strip with  anchored excluded-volume chains
 estimated for different edge lengths $L_y$
at $\phi=0$, $0.1$, $0.15$, $0.2$, and $0.25$. }
\end{figure}

Next, we investigate the edge line tension with various anchored
polymer densities for both ideal chains and excluded-volume chains.
A  strip of single-phase membrane with anchored polymers is used to
estimate the  edge tension  $\Gamma_{\rm {ed}}$ (see
Fig.~\ref{fig:snapstrip}). The  edge tension $\Gamma_{\rm {ed}}$ can
be calculated by \cite{Hay11,Briels04,Deserno08}
\begin{eqnarray}
\Gamma_{\rm {ed}} = \frac{\partial F}{2\partial L_y}= \Big\langle
\frac{P_{xx}+P_{zz}}{2}-P_{yy}\Big\rangle \frac{L_x L_z}{2},
\label{eq:lintenbio}
\end{eqnarray}
since  the total edge length is $2L_y$. The pressure
$P_{xx}=P_{zz}\approx 0$ for solvent-free simulation with a
negligibly low critical micelle concentration. We checked that the
edge tension is independent of the edge length for pure membranes as
well as for polymer-decorated membranes (see
Fig.~\ref{fig:nochange}).

\begin{figure}
  \includegraphics[width=8.cm]{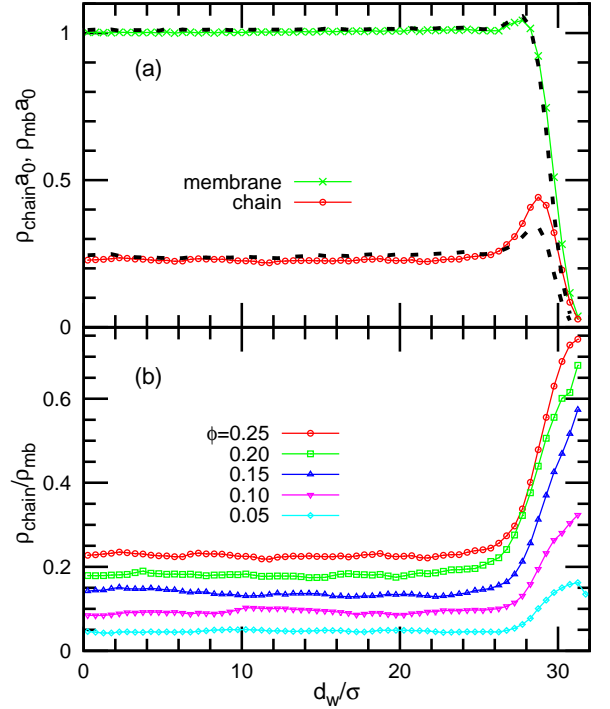}
  \caption{\label{fig:distribution}
Density distribution in the membrane strip. (a) Density of
polymer-anchored membrane particles $\rho_{\rm {chain}}$, and total
density $\rho_{\rm {mb}}$ at $\phi=0.25$. The solid lines with
symbols and dashed lines represent the data for the excluded-volume
chains, and ideal chains, respectively. (b) Density ratio $\rho_{\rm
{chain}}/\rho_{\rm {mb}}$ for the excluded-volume chains. The
distance $d_{\rm w}$ from the  center of the strip is taken in the
direction perpendicular to the edge. The membrane lengths are
$L_{\rm {st}}=60\sigma$ perpendicular to the edge and
$L_y=28.8\sigma$ along the edge. }
\end{figure}

Figure~\ref{fig:edgetension} shows that the edge tension
$\Gamma_{\rm {ed}}$ decreases with increasing polymer density
$\phi$. The reduction for excluded-volume chains is much larger than
that for ideal chains, similar to polymer effects on the bending
rigidity. The polymer chains prefer staying on the edge, since there
is more space to move so that they can gain entropy.
Figure~\ref{fig:distribution}(a) shows that the polymer density
distribution $\rho_{\rm {chain}}$ is nonuniform at the distance
$d_{\rm w}$ from the strip's central axis. High peaks of $\rho_{\rm
{chain}}$ are found close to the edges for both ideal chains and
excluded-volume chains, while the density  $\rho_{\rm {mb}}$ of all
membrane particles has only very small peak. The relative polymer
density  $\rho_{\rm {chain}}/\rho_{\rm {mb}}$ more rapidly increases
at the edges for larger mean density $\phi$ [see Fig.
~\ref{fig:distribution}(b)]. The mean polymer density $\phi_1$ at
the edges is calculated as an average $\langle\sum_{d_{\rm w} \ge
d_{\rm w}^{\rm {max}}} \rho_{\rm {chain}} /\sum_{d_{\rm w} \ge
d_{\rm w}^{\rm {max}}} \rho_{\rm {mb}}\rangle$ for the right region
of the peak ($d_{\rm w}^{\rm {max}}$) of $\rho_{\rm {mb}}$ in Fig.
\ref{fig:distribution}(a). The density difference from the mean
value  $\Delta \phi = \phi_1-\phi$ increases with increasing $\phi$
as shown in  Fig. \ref{fig:dphiphi}. The excluded volume chains
induce higher polymer concentration at the edges than the ideal
chains.

\subsection{Theoretical analysis}

Here we propose a mean field theory for the  edge line tension
induced by the anchored polymers  in the mushroom region. According
to the nonuniform polymer distribution on the membrane strip, we
divide the membrane into two regions, an edge (region 1) and middle
region (region 2). The polymer density is assumed to be uniform in
each region. The area fractions of the two regions are $n_1$ and
$n_2$ with $n_1+n_2=1$, and the polymer densities are $\phi_1$ and
$\phi_2$ with $\phi = n_1\phi_1 +n_2\phi_2$. The width of region 1
is considered the radius of gyration of polymer $R_{\rm g}$, so that
the area fraction is given by
\begin{equation}
  n_1 = \frac{2L_yR_{\rm g}}{N_{\rm {mb}}a_0}.
\end{equation}

The free energy of the membrane strip is written as
\begin{eqnarray}
\frac{F_{\rm {ed}}}{N_{\rm {mb}}k_{\rm B}T}&= &n_1\phi_1\ln\phi_1+n_1\left(1-\phi_1\right)\ln\left(1-\phi_1\right) \nonumber \\
&&+n_2\phi_2\ln\phi_2+n_2\left(1-\phi_2\right)\ln\left(1-\phi_2\right)\nonumber
\\ &&-n_1\phi_1\Delta S+f_0,
\end{eqnarray}
where $f_0$ is the free energy contribution of the membrane without
polymer anchoring. The first four terms are the mixing entropy for
regions 1 and 2. When a polymer chain moves from the middle region
to the open edges, it gains the conformational entropy $\Delta S$.

\begin{figure}
  \includegraphics[width=8.cm]{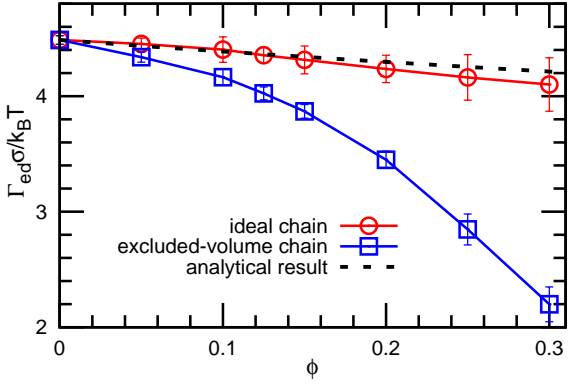}
  \caption{\label{fig:edgetension}
Polymer density dependence $\phi$ of the edge line tension for ideal
and excluded-volume chains. The dashed line represents our
theoretical prediction by Eq.~\eqref{eq:ged2}. }
\end{figure}

The partition function of a single anchored polymer chain is
expressed as $Z_{\rm p}=q^{N_{\rm p}}W$, where $q$ is the number of
the nearest neighbors in the lattice model ($q=6$ in a cubic
lattice). The restricted weight of a polymer anchored on the flat
membrane is $W_{\rm {hs}}={\rm {erf}}\left[\frac{\sqrt{q}l_{\rm
{an}}}{2R_{\rm {end}}}\right]$, where erf$(x)$ is the error function
and $l_{\rm {an}}$ is the anchor length \cite{Lip95,Lip96}. On the
other hand, the free end of a polymer anchored on the edge can also
move in the other half space, and has a larger value of weight
$W_{\rm {ed}}$. We numerically counted the weights $W_{\rm {ed}}$
and $W_{\rm {hs}}$ in a cubic lattice. The ratio $W_{\rm
{ed}}/W_{\rm {hs}}$ increases with increasing $N_{\rm p}$, and
$W_{\rm {ed}}/W_{\rm {hs}} \simeq 2$ for $N_{\rm p}=10$. Thus, the
excess entropy is estimated as $\Delta S=\ln\left(W_{\rm
{ed}}/W_{\rm {hs}}\right)\simeq \ln2$ for our simulation condition.

\begin{figure}
  \includegraphics[width=8.cm]{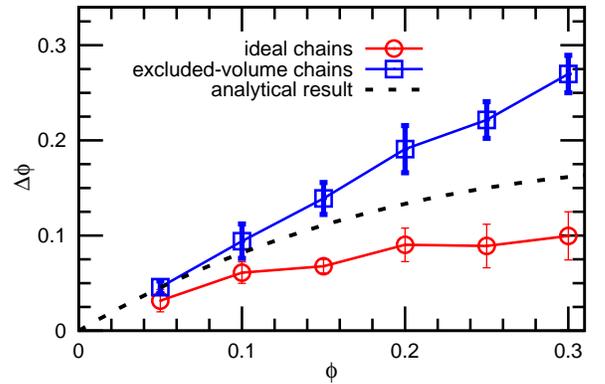}
  \caption{\label{fig:dphiphi}
Excess polymer density $\Delta \phi\equiv \phi_1-\phi$ at the
membrane edge as a function of the mean polymer density $\phi$. The
solid lines with circles and squares  represent our simulation
results for  the ideal chains and excluded volume chains,
respectively. The dashed line represents our theoretical prediction
for the ideal chains by Eq.~\eqref{eq:dphi}. }
\end{figure}

Using minimization of $F_{\rm {ed}}$, the polymer density $\phi_1$
in the edge region is analytically derived as
\begin{eqnarray}\label{eq:dphi}
\phi_1 &=& \frac{2Q\phi}{s+ \sqrt{ s^2 -4Q(Q-1)\phi n_1} } \\
\nonumber
 &=& \frac{Q\phi}{1+(Q-1)\phi}\Bigg( 1 - \frac{(Q-1)(1-\phi )}{\{1+(Q-1)\phi\}^2}n_1 \Bigg) + O(n_1^2),
\end{eqnarray}
where $Q=\exp(\Delta S)$ and $s=1+(Q-1)(\phi+n_1)$. At $Q=2$ and
$n_1 \ll 1$, the density difference is simply $\Delta \phi \equiv
\phi_1-\phi=\phi(1-\phi)/(1+\phi)$, which agrees well with the
simulation results (see Fig. \ref{fig:dphiphi}).

The  edge tension is derived as $\Gamma_{\rm {ed}}= \partial F_{\rm
{ed}}/\partial L_{\rm {ed}}$, where $L_{\rm {ed}}$ is the total edge
length $L_{\rm {ed}}=2L_y$. Thus, the polymer-induced  edge tension
$\Delta \Gamma_{\rm {ed}}$ is given by
\begin{eqnarray}
\frac{\Delta\Gamma_{\rm {ed}}a_0}{R_{\rm g}k_{\rm B}T} &=&
\ln(1-n_1) +
  \phi_1\ln\frac{\phi_1}{Q( \phi - \phi_1n_1 )}  \\  \nonumber
& &+  (1-\phi_1 )\ln\frac{1-\phi_1}{1-\phi - (1-\phi_1)n_1 }  .
\end{eqnarray}
At  $Q=2$, the Taylor expansion gives
\begin{equation}\label{eq:ged2}
 \frac{\Delta\Gamma_{\rm {ed}}a_0}{R_{\rm g}k_{\rm B}T} =  -\ln(1+\phi) + \frac{\phi (1-\phi)}{(1+\phi)^2}n_1  + O(n_1^2).
\end{equation}
Thus, the edge tension $\Gamma_{\rm {ed}}$ decreases with increasing
$\phi$ and is independent of the edge length $L_y$ for $n_1 \ll 1$.
Figure~\ref{fig:edgetension} shows the comparison of edge tensions
between our simulation and the theoretical results for ideal chains;
The agreement is excellent. As the membrane strip becomes narrower
($n_1$ increases), the polymer effect on the edge tension
$\Gamma_{\rm {ed}}$ is reduced by the loss of mixing entropy in
region 2, and  $\Gamma_{\rm {ed}}$ increases with increasing edge
length $L_y$.

For the excluded-volume polymer chains, the membrane cannot be
simply divided into two regions because of the repulsive interaction
between polymer chains. The effects of the membrane edges may be
considered as an increase in the average volume for each chain. For
a flat membrane without edges, the volume per chain is given by
$V_{\rm {pf}}=2R_{\rm {end}}L_x L_y/N_{\rm {chain}}$. The membrane
strip has an additional space $\pi R_{\rm {end}}^2 L_y$ around the
edges so that the polymer volume becomes
 $V_{\rm {pe}}=V_{\rm {pf}} + \pi R_{\rm {end}}^2 L_y/N_{\rm {chain}}$.
Thus, the polymer chains gain additional conformational entropy not
only at the edges but in the middle of the strip.

\section{Membrane domains with anchored polymers}
\label{sec:dom}

\begin{figure}
  \includegraphics{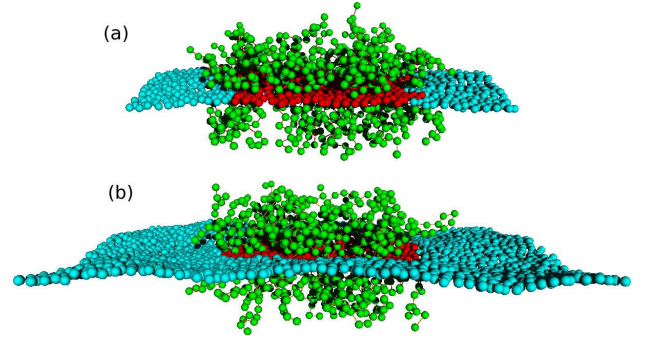}
  \caption{\label{fig:snap1} Snapshots of (a) striped and (b) circular domains
with  anchored excluded-volume chains in  planar membranes at
$N_{\rm A}=400$ and $\phi=0.3$. Type A and B membrane particles are
displayed in red and blue, respectively. }
\end{figure}

In this section, we focus on the effects of anchored polymers on
membrane phase separation. First, in Sec. \ref{sec:ltdm} we estimate
the line tension of polymer-anchored membrane domains, and then in
Sec. \ref{sec:dms} we investigate the polymer effects on domain
separation and domain shape transformation. Here, we investigate
only the membranes with excluded-volume chains, since the effects of
the ideal chains are considered to be very small. As described in
Sec. \ref{sec:scv}, polymers can induce an effective spontaneous
curvature in the membrane. In order to  diminish the influence of
the induced spontaneous curvature, we symmetrically anchor
 polymer chains on both sides of the membrane
as shown in Fig. \ref{fig:snap1}. Half of the chains ($N_{\rm
{chain}}/2$) are  anchored on the upper (lower) side of the
membrane, and each chain is anchored on one membrane particle. Then,
the net curvature effects induced on both sides of the membrane
 cancel each other out.

\subsection{Interfacial line tension between two membrane domains}
\label{sec:ltdm}

\begin{figure}[tpc]
  \includegraphics[width=8.cm]{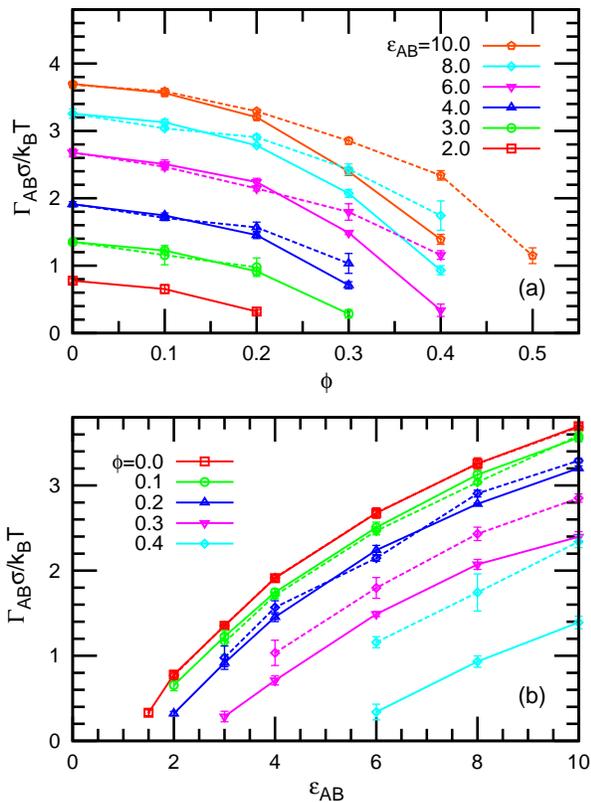}
  \caption{\label{fig:interfacetension}
Interfacial line tension $\Gamma_{\rm {AB}}$ between membrane
domains as a function of (a) $\phi$ and (b) $\varepsilon_{\rm
{AB}}$. The solid and dashed lines represent  $\Gamma_{\rm {AB}}$
estimated from the striped domain and the circular domain,
respectively. }
\end{figure}

The line tension $\Gamma_{\rm {AB}}$ between the type A and B
domains is estimated by two methods using a striped domain and a
circular domain. For the  striped domain shown in
Fig.~\ref{fig:snap1}$(a)$, the line tension is calculated by
\begin{eqnarray}
\Gamma_{\rm {AB}} = \langle P_{xx}-P_{yy}\rangle L_x L_z/2.
\label{eq:lintensmp}
\end{eqnarray}
The obtained line tension for tensionless membranes is shown by
solid lines in Fig.~\ref{fig:interfacetension}. We ensured that
$\Gamma_{\rm {AB}}$ is independent of the boundary length $L_y$ for
$24<L_y/\sigma<48$ (data not shown). The line tension $\Gamma_{\rm
{AB}}$ decreases with increasing $\phi$, while $\Gamma_{\rm {AB}}$
increases with increasing $\varepsilon_{\rm {AB}}$. Thus, the same
value of $\Gamma_{\rm {AB}}$ can be obtained for different polymer
densities $\phi$ by adjusting $\varepsilon_{\rm {AB}}$.

Before investigating polymer effects on the domain shapes in the
next subsection,
 we also estimate  $\Gamma_{\rm {AB}}$ from the circular domain
 shown in Fig.~\ref{fig:snap1}$(b)$.
The  line tension  $\Gamma_{\rm {AB}}$ is calculated by the 2D
Laplace pressure, \cite{Briels04,nog12}
\begin{eqnarray}
\Gamma_{\rm {AB}} = \bar{R} \Delta\gamma, \label{eq:raftline}
\end{eqnarray}
where $\bar{R}$ is the average radius of the domain, and
$\Delta\gamma$ is the difference of surface tension between the type
A and B domains: $\Delta\gamma=\gamma^{\rm {in}}-\gamma^{\rm
{out}}$, where $\gamma^{\rm {in}}$ is the surface tension of the
inner (type A) domain and $\gamma^{\rm {out}}$ is that of the outer
(type B) domain. Both of them can be estimated by the pressure
tensors of the local regions
\begin{eqnarray}
\gamma^{\rm {\alpha}} = \langle P^{\rm {\alpha}}_{zz}-(P^{\rm
{\alpha}}_{xx}+P^{\rm {\alpha}}_{yy})/2\rangle L_z,
\label{eq:raftsurf}
\end{eqnarray}
where $\alpha$ represents ``in'' or ``out''; $P^{\rm
{\alpha}}_{xx}$, $P^{\rm {\alpha}}_{yy}$, and $P^{\rm
{\alpha}}_{zz}$ are the diagonal components of the pressure tensors
calculated in the local membrane regions. The outer surface tension
$\gamma^{\rm {out}}$ can also be calculated from the pressure
tensors for the whole area.

To estimate $\gamma^{\rm {in}}$ and $\gamma^{\rm {out}}$, we extract
the inner and outer regions as follows. First, domains of type A
particles are calculated. The particles are considered to belong to
the same cluster (domain) when their distance is less than $r_{\rm
{cut}}=2.1\sigma$. Then the radius $\bar{R}$ of the largest domain
is calculated. Type A particles contacting type B particles (closer
than $r_{\rm {cut}}$) are considered domain boundary particles. The
number of boundary particles is $N_{\rm {bd}}$. In the largest
domain, the distance of the domain particles from the center ${\bf
r}_{\rm G}$ of the domain is averaged by $R_{\rm A}= (1/N_{\rm
{bd}})\sum |{\bf r}-{\bf r}_{\rm G}|$. For the mean radius of the
domain boundary, the half boundary width $\sqrt{a_0}/2=0.6\sigma$ is
added so that $\bar{R}=R_{\rm A}+0.6\sigma$. Then, the maximum
fluctuation amplitude $\Delta R$ around $\bar{R}$ is calculated. The
surface tension $\gamma^{\rm {in}}$ is estimated within the area
inside the circular region with  radius $\bar{R}-\Delta
R-0.5\sigma$, while $\gamma^{\rm {out}}$ is estimated within the
area outside the circular region with radius $\bar{R}+\Delta
R+0.5\sigma$. Note that a few type B particles can enter the type A
domain at small $\Gamma_{AB}\sigma/k_{\rm B}T \sim 1$ so the type A
particles neighboring these isolated particles are not taken into
account for estimation of $\bar{R}$ and $\Delta R$.

The  line tension estimated from the 2D Laplace pressure is shown by
dashed lines in Fig.~\ref{fig:interfacetension}. For the pure
membrane, the obtained values agree with those from the membrane
strip very well. However, they are slightly larger for the
polymer-anchored membranes. This deviation is likely caused by the
relative larger boundary region of the circular domain than the
striped domain. It is a similar dependence obtained for the membrane
edges (see Eq. \eqref{eq:ged2}).

The phase behavior of the pure membranes ($\varepsilon_{\rm
{AB}}=0$) belongs to the universality class of the 2D Ising
model~\cite{lipo92}. For the polymer-anchored membranes, however,
the line tension dependence on the boundary curvature is not
explained by the universality class. Thus, the polymer effects
cannot be treated as an effective potential between neighboring
membrane particles.

\begin{figure}
  \includegraphics[width=8.cm]{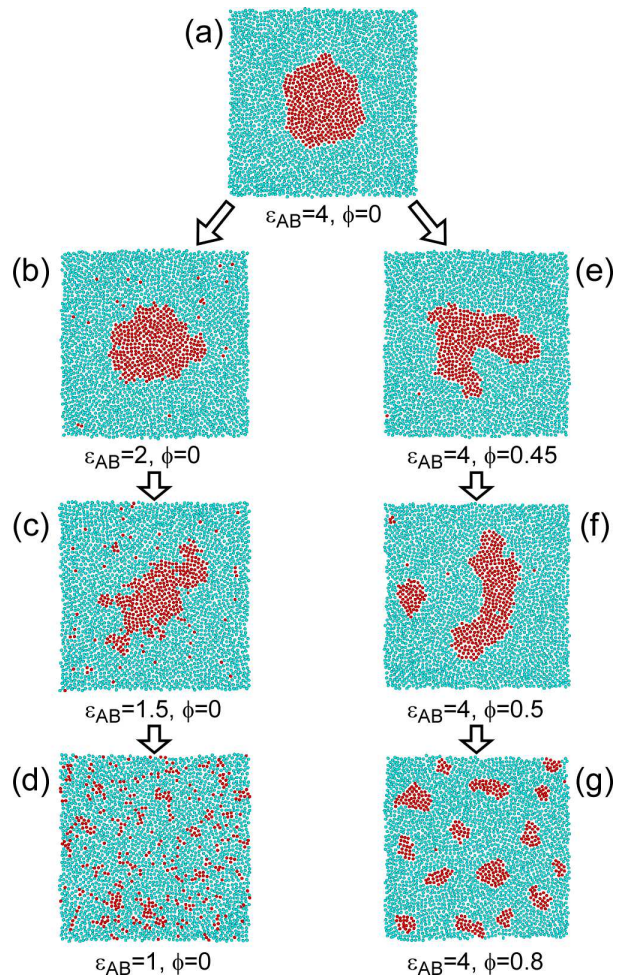}
  \caption{\label{fig:snapsep}
Sequential snapshots of  domain separations as (left from (a) to
(d)) $\varepsilon_{\rm {AB}}$
 decreases at $\phi=0$ or (right from
(a) to (g)) as the polymer density $\phi$ increases at
$\varepsilon_{\rm {AB}}=4$. Red and blue particles represent the A
and B type membrane particles, respectively. To show microdomain
separation and shape transformation clearly, polymer particles are
not displayed. }
\end{figure}

\subsection{Domain separation and microdomain formation}
\label{sec:dms}

\begin{figure}
  \includegraphics[width=8.cm]{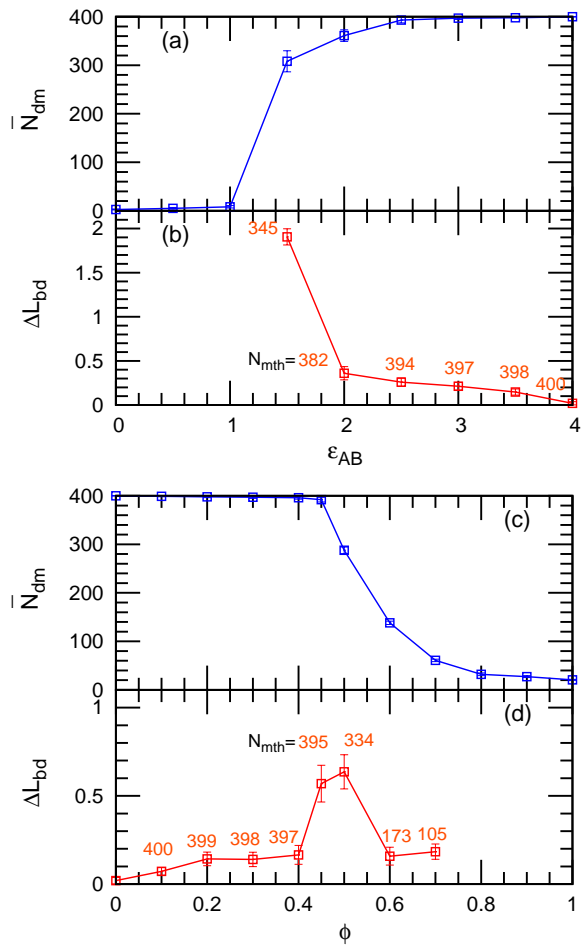}
  \caption{\label{fig:clustersize}
Domain shape changes and domain separation as (a,b)
$\varepsilon_{\rm {AB}}$ decreases at $\phi=0$ and (c,d) $\phi$
increases at $\varepsilon_{\rm {AB}}=4$. (a,c) The average cluster
size $\bar{N}_{\rm {dm}}$ of the mother (largest) domains and (b,d)
the reduced excess domain length $\Delta L_{\rm {bd}}$ of the mother
domains. The mean number $N_{\rm {mth}}$ of the membrane particles
in the mother domains at each stage is shown in light red color. }
\end{figure}

To clarify the anchored polymer effects, we compare the shape
changes of the type A membrane domains with increasing $\phi$ and
with decreasing $\varepsilon_{\rm {AB}}$. In both cases, the
interfacial line tension $\Gamma_{\rm {AB}}$ decreases and the low
line tension leads to the breakup of domains. However, the resultant
states are quite different as shown in Fig. \ref{fig:snapsep}. As
the repulsive interaction between the type A and B particles is
reduced with decreasing $\varepsilon_{\rm {AB}}$, the obtained phase
behavior is similar to that of typical binary fluids. At
$\Gamma_{\rm {AB}}\sigma \simeq k_{\rm B}T$ ($\varepsilon_{\rm
{AB}}=2$), the domain boundary undergoes large fluctuation and a few
(type A or B) particles leave their domain to dissolve in the other
domain. As $\varepsilon_{\rm {AB}}$ decreases further, the domain
breaks up into small domains, and finally the two types of particles
are completely mixed.

On the other hand, the anchored polymers induce formation of small
stable domains (called microdomains) instead of a mixing state,
although it can reduce the line tension to $\Gamma_{\rm {AB}}\sigma
\lesssim k_{\rm B}T$ (see Fig.~\ref{fig:snapsep}(g)). At $\phi \leq
0.45$, the type A domain remains as one domain but exhibits an
elongated shape at $\phi=0.45$. At $\phi \geq 0.5$, it starts
separating into microdomains. Note that the membrane is considered
in a mixed state even at $\phi=0.45$, if $\Gamma_{\rm {AB}}$ for the
straight boundary is extrapolated (see
Fig.~\ref{fig:interfacetension}).

In contrast to the reduction in $\varepsilon_{\rm {AB}}$, the
boundary of the elongated domain is rather smooth (compare snapshots
in Figs. \ref{fig:snapsep}(c) and (f)). We confirmed that these
small domains are also formed from random distribution of initial
states. Thus, it is a thermodynamically stable state.

Let us discuss the effects of the polymer anchoring on the domain
formation. First, we remind that the polymer beads have only
repulsive interactions with the other beads and membrane particles
except for the membrane-anchored head particles.

The polymer effects seem suppressed for shorter lengths than the
polymer size $\sim R_{\rm {end}} = 4\sigma$. A smaller boundary
undulation than the polymer size does not yield additional space for
the polymer brush. A similar suppression in the short length scale
was reported on the bending rigidity induced by the polymer
anchoring \cite{Auth05}. When the domain size is comparable to the
polymer length, most of the particles already stay at the domain
boundary, so that
 an additional increase
in the boundary length likely yields much less gain in the average
volume per polymer and the polymer conformational entropy. As
explained in Sec. \ref{sec:ltdm}, the line tension of the circular
domain is larger than the straight boundary. For the smaller
domains, this difference would be enhanced, although the domains are
too small for direct estimation of $\Gamma_{\rm {AB}}$ by Laplace's
law.

To investigate the changes of domains in greater detail, we
calculate the mean cluster size $\bar{N}_{\rm {dm}}$ and a reduced
excess domain length $\Delta L_{\rm {bd}}$. The cluster size
$\bar{N}_{\rm {dm}}$ is defined as
\begin{equation}
\bar{N}_{\rm {dm}} =\frac{\sum^{N_{\rm A}}_{i_{\rm c}=1}n_ii_{\rm
c}^2}{\sum^{N_{\rm A}}_{i_{\rm c}=1}n_ii_{\rm c}},
\label{eq:avdomain}
\end{equation}
where $n_i$ is the number of clusters with size $i_{\rm c}$. The
reduced excess domain length for the mother (largest) domains
$\Delta L_{\rm {bd}}$ is defined as
\begin{equation}
\Delta L_{\rm {bd}} =\frac{L_{\rm {bd}}}{2\sqrt{\pi A_{\rm
{dm}}}}-1, \label{eq:reducelength}
\end{equation}
where $L_{\rm {bd}} = N_{\rm {bd}}\sqrt{a_0}$ is the boundary length
of the mother domains and
  $A_{\rm {dm}}=N_{\rm A}a_0$ is the domain area.
The length $L_{\rm {bd}}$ is normalized by the length of a circular
domain $2\sqrt{\pi A_{\rm {dm}}}$ so that $\Delta L_{\rm {bd}} =0$
for the circular domain.

Figure~\ref{fig:clustersize} shows the development of $\bar{N}_{\rm
{dm}}$ and  $\Delta L_{\rm {bd}}$. In the $\varepsilon_{\rm {AB}}$
reduction, the transition to the mixing state occurs sharply between
$\varepsilon_{\rm {AB}}=1.5$ and $1$. However, for polymer
  anchoring, a gradual decrease in $\bar{N}_{\rm {dm}}$ represents the
formation of microdomains (see Fig.~\ref{fig:clustersize}(c)).
Around the transition points,
 $\Delta L_{\rm {bd}}$ is increased less by polymer anchoring than by
lowering $\varepsilon_{\rm {AB}}$, while both domains are similarly
elongated (see Fig.~\ref{fig:snapsep}). This difference is caused by
the weaker undulation of polymer-decorated domain boundaries.

\begin{figure}
  \includegraphics[width=8.cm]{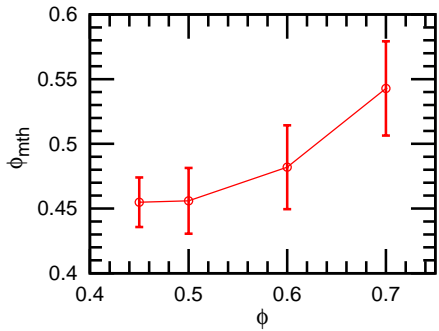}
  \caption{\label{fig:motheranchor}
Mean polymer density $\phi$ dependence of the number fraction
$\phi_{\rm {mth}}$ of polymer chain anchors on the mother (largest)
domain. }
\end{figure}

We calculated the fraction of polymer chain anchors on the mother
domain $\phi_{\rm {mth}}$ after the microdomain separation (see
Fig.~\ref{fig:motheranchor}). Interestingly, it is lower than the
initial density $\phi$. Thus,  detached small domains have higher
polymer densities than their mother domain. This is caused
thermodynamically by the entropy gain of polymers anchored on small
domains and also kinetically by a higher density at the domain
boundary.

\section{Summary and discussion}

We have systematically studied the entropic effects of anchored
polymers on various types of mechanical and interfacial properties
of biomembranes using particle-based membrane simulations. First, we
reconfirm the previous theoretical predictions for spontaneous
curvature and bending rigidity by simulating cylindrical membranes.
They increase with the anchored polymer density $\phi$ linearly in
the mushroom region, but sharply increase in the brush region.

Second, we investigated the polymer anchoring effects on the edge
line tension for ideal and excluded-volume chains. It is revealed
that polymer  anchoring significantly reduces the edge tension. For
ideal polymer chains, it is also investigated by a mean field
theory. It is clarified that the entropy gain of polymer
conformation at the membrane edge reduces the edge tension.
Experimentally, it is known that polymer   anchoring induces the
formation of large vesicles \cite{bres12} and spherical or discoidal
micelles \cite{john03}. Since the ratio between the edge tension and
the bending rigidity determines the vesicle radius $R_{\rm {ves}}$
formed by the membrane disks as $R_{\rm {ves}} \sim
(2\kappa+\bar{\kappa})/\Gamma_{\rm {ed}}$, the reduction in the edge
tension increases the vesicle radius. Our results are consistent
with these experimental observations.

Finally, we investigated the polymer  anchoring effects on
 phase separation in membranes for excluded-volume chains.
The line tension of the domain boundary is reduced by  anchoring
polymers. It is found that densely
 anchored polymers can stabilize microdomains, whereas large domains are
unstable. Although we did not investigate polymer length dependence
here, it is expected that the domain size can be controlled by the
polymer length. In living cells, lipid rafts contain a large amount
of glycosphingolipids \cite{Vereb03,Korade,Pike}. Our simulation
results suggest that the entropic effects of glycosphingolipids may
play a significant role in stabilizing microdomains $\lesssim 100$
nm. At a moderate polymer density, elongated shapes of membrane
domains are obtained. In lipid membranes with PEG-conjugated
cholesterol, the domain shapes depend on the  anchored polymer
density $\phi_{\rm {PEG}}$: at a high $\phi_{\rm {PEG}}$, small
domains are scattered, while at a slightly lower $\phi_{\rm {PEG}}$,
small elongated domains are connected with each other to form a
network~\cite{Miho12}. The elongated domains in our simulations may
form a network, if much larger domains are simulated. Further study
is needed to clarify the polymer-anchoring effects on large-scale
domain patterns.

Our present study highlights entropic effects of anchored polymers
on the microdomain formation via the reduction in domain boundary
tension on quasi-2D biomembranes. It is well known that high line
tension can induce budding of membranes \cite{lipo03,baga09,lipo92}.
Nonzero spontaneous curvature induced by proteins and anchored
polymers can lead to various liposome shapes,
 such as tube formation and pearling \cite{baum11,phil09,shny09,tsaf01,akiy03,guo09}.
Shape transformation of vesicles induced by polymer-decorated
domains is an interesting topic for further studies.

\section{Acknowledgments}
We would like to thank G. Gompper, P. A. Pincus, T. Auth, T.
Taniguchi, V. N. Manoharan, and S. Komura for informative
discussions. This study is partially supported by KAKENHI (25400425)
from the Ministry of Education, Culture, Sports, Science, and
Technology (MEXT) of Japan. HW is supported by a MEXT scholarship.


\begin{thebibliography}{99}

\bibitem{Singer72}
S.~J. Singer and G.~L. Nicolson, \emph{Science}, \textbf{175}, 720
(1972).

\bibitem{Simons97}
K.~Simons and E.~Ikonen, \emph{Nature}, \textbf{387}, 569 (1997).

\bibitem{Ikonen01}
E.~Ikonen, \emph{Curr. Opin. Cell Biol.}, \textbf{13}, 470 (2001).

\bibitem{Vereb03}
G.~Vereb, J.~Sz{\"o}ll{\H{o}}si, J.~Matko, P.~Nagy, T.~Farkas,
L.~Vigh, L.~Matyus, T.~A. Waldmann and S.~Damjanovich, \emph{Proc.
Natl. Acad. Sci. USA}, \textbf{100}, 8053  (2003).

\bibitem{Korade}
Z.~Korade and A.~K. Kenworthy, \emph{Neuropharmacology},
\textbf{55}, 1265 (2008).

\bibitem{Pike}
L.~J. Pike, \emph{J. Lipid Res.}, \textbf{50}, S323 (2009).

\bibitem{hone09}
A.~R. Honerkamp-Smith, S.~L. Veatch and S.~L. Keller,
\emph{Biochim.\ Biophys.\ Acta}, \textbf{1788}, 53 (2009).

\bibitem{lipo03}
R.~Lipowsky and R.~Dimova, \emph{J.\ Phys.:\ Condens. Matter},
  \textbf{15}, S31 (2003).

\bibitem{baga09}
L.~Bagatolli and P.~B.~S. Kumar, \emph{Soft\ Matter}, \textbf{5},
  3234 (2009).

\bibitem{Keller02}
S.~L. Veatch and S.~L. Keller, \emph{Phys. Rev. Lett.}, \textbf{89},
268101 (2002).

\bibitem{Tobias03}
T.~Baumgart, S.~T. Hess and W.~W. Webb, \emph{Nature}, \textbf{425},
  821 (2003).

\bibitem{Hammond05}
A.~T. Hammond, F.~A. Heberle, T.~Baumgart, D.~Holowka, B.~Baird and
G.~W. Feigenson, \emph{Proc. Natl. Acad. Sci. USA}, \textbf{102},
  6320 (2005).

\bibitem{Kuzmin05}
P.~I. Kuzmin, S.~A. Akimov, Y.~A. Chizmadzhev, J.~Zimmerberg and
F.~S. Cohen, \emph{Biophys. J.}, \textbf{88}, 1120 (2005).

\bibitem{yana08}
M.~Yanagisawa, M.~Imai and T.~Taniguchi, \emph{Phys.\ Rev.\ Lett.},
\textbf{100}, 148102 (2008).


\bibitem{Schick09}
G.~G. Putzel and M.~Schick, \emph{Biophys. J.}, \textbf{96},
  4935 (2009).


\bibitem{mcco88}
H.~M. McConnell and V.~T. Moy, \emph{J. Phys. Chem.}, \textbf{92},
4520 (1988).

\bibitem{HaoWu09}
H.~Wu and Z.~C. Tu, \emph{J. Chem. Phys.}, \textbf{130}, (2009).

\bibitem{iwam10}
M.~Iwamoto, F.~Liu and Z.-c. Ou-Yang, \emph{EPL}, \textbf{91}, 16004
(2010).

\bibitem{Miho12}
M.~Yanagisawa, N.~Shimokawa, M.~Ichikawa and K.~Yoshikawa,
\emph{Soft Matter}, \textbf{8}, 488 (2012).

\bibitem{Lip95}
R.~Lipowsky, \emph{Europhys. Lett.}, \textbf{30}, 197 (1995).

\bibitem{Lip96}
C.~Hiergeist and R.~Lipowsky, \emph{J. Phys. II France}, \textbf{6},
1465 (1996).

\bibitem{brei00}
M.~Breidenich, P.~R. Netz, and R.~Lipowsky, \emph{Europhys. Lett.},
\textbf{49}, 431 (2000).

\bibitem{Auth03}
T.~Auth and G.~Gompper, \emph{Phys. Rev. E}, \textbf{68}, 051801
(2003).

\bibitem{Auth05}
T.~Auth and G.~Gompper, \emph{Phys. Rev. E}, \textbf{72}, 031904
(2005).

\bibitem{wern10}
M.~Werner and J.-U. Sommer, \emph{Eur. Phys. J. E}, \textbf{31}, 383
(2010).

\bibitem{Marsh03}
D.~Marsh, R.~Bartucci, and L.~Sportelli, \emph{Biochim. Biophys.
Acta}, \textbf{1615}, 33 (2003).

\bibitem{Evans97}
E.~Evans and W.~Rawicz, \emph{Phys. Rev. Lett.}, \textbf{79}, 2379
(1997).

\bibitem{lasi94}
D.~D. Lasic, \emph{Angew. Chem. Int. Ed. Engl.}, \textbf{33}, 1685
(1994).

\bibitem{hoff08}
A.~S. Hoffman, \emph{J. Control. Release}, \textbf{132}, 153 (2008).

\bibitem{helf74}
W.~Helfrich, \emph{Phys. Lett. A}, \textbf{50}, 115 (1974).

\bibitem{from83}
P.~Fromherz, \emph{Chem.\ Phys.\ Lett.}, \textbf{94}, 259 (1983).

\bibitem{kale89}
E.~W. Kaler, A.~K. Murthy, B.~E. Rodriguez, and J.~A.~N.
Zasadzinski,\emph{Science}, \textbf{245}, 1371 (1989).

\bibitem{weis05}
T.~M. Weiss, T.~Narayanan, C.~Wolf, M.~Gradzielski, P.~Panine,
S.~Finet, and W.~I. Helsby, \emph{Phys.\ Rev.\ Lett.}, \textbf{94},
038303 (2005).

\bibitem{leng02}
J.~Leng, S.~U. Egelhaaf, and M.~E. Cates, \emph{Europhys.\ Lett.},
\textbf{59}, 311 (2002).

\bibitem{made11}
D.~Madenci, A.~Salonen, P.~Schurtenberger, J.~S. Pedersen, and S.~U.
Egelhaaf, \emph{Phys.\ Chem.\ Chem.\ Phys.}, \textbf{13}, 3171
(2011).

\bibitem{brys05}
K.~Bryskhe, S.~Bulut, and U.~Olsson, \emph{J.\ Phys.\ Chem. B},
\textbf{109}, 9265 (2005).

\bibitem{nogu06a}
H.~Noguchi and G.~Gompper, \emph{J.\ Chem.\ Phys.}, \textbf{125},
164908 (2006).

\bibitem{nogu13}
H.~Noguchi, \emph{J.\ Chem.\ Phys.}, \textbf{138}, 024907 (2013).

\bibitem{bres12}
K.~Bressel, M.~Muthig, S.~Prevost, J.~Gummel, T.~Narayanan, and
M.~Gradzielski, \emph{ACS Nano}, \textbf{6}, 5858 (2012).

\bibitem{scha02}
A.~Schalchli-Plaszczynski and L.~Auvray, \emph{Eur. Phys. J. E},
\textbf{7}, 339 (2002).

\bibitem{john03}
M.~Johnsson and K.~Edwards, \emph{Biophys. J.}, \textbf{85}, 3839
(2003).

\bibitem{nog09}
H.~Noguchi, \emph{J. Phys. Soc. Jpn.}, \textbf{78}, 041007 (2009).

\bibitem{nog06}
H.~Noguchi and G.~Gompper, \emph{Phys. Rev. E}, \textbf{73}, 021903
(2006).

\bibitem{HaoWu13}
H.~Wu and H.~Noguchi, \emph{AIP Conf. Proc.}, \textbf{1518}, 649
(2013).

\bibitem{nog12}
H.~Noguchi, \emph{Soft Matter}, \textbf{8}, 8926 (2012).

\bibitem{alle87}
M.~P. Allen and D.~J. Tildesley, \emph{Computer Simulation of
Liquids}, Clarendon Press, Oxford (1987).

\bibitem{Hay11}
H.~Shiba and H.~Noguchi, \emph{Phys. Rev. E}, \textbf{84}, 031926
(2011).

\bibitem{nogu11a}
H.~Noguchi, \emph{Phys. Rev. E}, \textbf{83}, 061919 (2011).

\bibitem{harm06}
V.~A. Harmandaris and M.~Deserno, \emph{J.\ Chem.\ Phys.},
\textbf{125}, 204905 (2006).

\bibitem{Briels04}
T.~V. Tolpekina, W.~K. den Otter, and W.~J. Briels, \emph{J. Chem.
Phys.}, \textbf{121}, 8014 (2004).

\bibitem{Deserno08}
B.~J. Reynwar and M.~Deserno, \emph{Biointerphases}, \textbf{3},
FA117 (2008).

\bibitem{lipo92}
R.~Lipowsky, \emph{J.\ Phys.\ II France}, \textbf{2}, 1825 (1992).

\bibitem{phil09}
R.~Phillips, T.~Ursell, P.~Wiggins, and P.~Sens, \emph{Nature},
\textbf{459}, 379 (2009).

\bibitem{shny09}
A.~V. Shnyrova, V.~A. Frolov, and J.~Zimmerberg, \emph{Curr.\
Biology}, \textbf{19}, R772 (2009).

\bibitem{tsaf01}
I.~Tsafrir, D.~Sagi, T.~Arzi, M.-A. Guedeau-Boudeville, V.~Frette,
D.~Kandel, and J.~Stavans, \emph{Phys. Rev. Lett.}, \textbf{86},
1138 (2001).

\bibitem{akiy03}
K.~Akiyoshi, A.~Itaya, S.~M. Nomura, N.~Ono, and K.~Yoshikawa,
\emph{FEBS Lett.}, \textbf{534}, 33 (2003).

\bibitem{guo09}
K.~Guo, J.~Wang, F.~Qiu, H.~Zhang, and Y.~Yang, \emph{Soft Matter},
\textbf{5}, 1646 (2009).

\bibitem{baum11}
T.~Baumgart, B.~R. Capraro, C.~Zhu, and S.~L. Das, \emph{Annu. Rev.
Phys. Chem.}, \textbf{62}, 483 (2011).



\end{thebibliography}
\end{document}